# AUTOMATED REAL-TIME TESTING (ARTT) FOR EMBEDDED CONTROL SYSTEMS (ECS)

Jon Hawkins, Group Leader, Interlock System & Instrumentation Group

Haung V. Nguyen, Engineering Assistant, Interlock System & Instrumentation Group,

Reginald B. Howard, Test Engineer Consultant

Advanced Photon Source, Bldg 401, L1111, 9700 S. Cass Ave, Argonne, IL 60439

Abstract

Developing real-time automated test systems for embedded control systems has been a real problem. Some engineers and scientists have used customized software and hardware as a solution, which can be very expensive and time consuming to develop. We have discovered how to integrate a suite of commercially available off-the-shelf software tools and hardware to develop a scalable test platform that is capable of performing complete black-box testing for a dual-channel real-time Embedded-PLC-based control system (www.aps.anl.gov). We will discuss how the Vali/Test Pro testing methodology was implemented to structure testing for a personnel safety system with large quantities of requirements and test cases.

This work was supported by the U.S. Department of Energy, Basic Energy Sciences, under Contract No. W-31-109-Eng-38.

## 1.0 INTRODUCTION

Many of today's automated real-time testing systems for embedded systems were developed using expensive custom hardware and software. In this article we describe how to use commercially available off-the-shelf hardware and software to design and develop an automated real-time test system for Embedded Programmable Logic Controller (PLC) Based Control Systems. Our system development began with the implementation of the VALI/TEST Pro testing methodology as a means for structuring the testing. Using this methodology, we were able to decompose system requirement documents for a Personnel Safety System (PSS) into its high, intermediate, and detail level requirements. Next, the validation procedures for the PSS system were decomposed into testing units called builds, test runs, and test cases. To measure the PSS system's test coverage, three levels of system requirements were mapped to their respective unit level of test using a specially constructed validation matrix that was designed to handle over 150 test cases and requirements. All of the above work led to the development of an Automated Real-Time Test System (ARTTS) that is capable of performing complete black box testing in real-time for Embedded PLC Based Control Systems. Also note, that the PSS system under test and mentioned in this paper is located at the Advanced Photon Source (APS) at Argonne National Laboratory Basic Energy Science Facility in Argonne, Illinois (www.aps.anl.gov).

## 2.0 PSS SYSTEM OPERATION

In this section, we explain the theory of operation of a Personnel Safety System (PSS) for which the prototype ARTTS System was designed to test.

### 2.1  PSS System Description

Consider figure 1. It depicts a typical configuration of "Station A", which is a major component of the overall PSS system. Note that the overall function of a PSS system is by definition, a highly reliable, fail-safe, redundant, stand-alone system that closely monitor and control personnel access into potentially hazardous Experimental Stations. It is also responsible for reducing hazards to mitigate harm to personnel against direct X-ray radiation from the Advanced Photon Source.

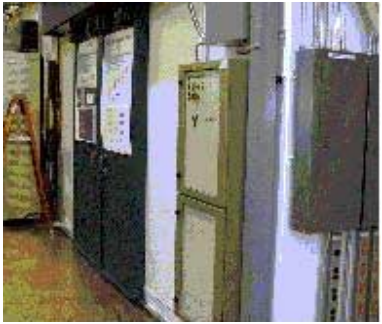

Fig. 1 Personnel Safety System (PSS)

## 2.2 PSS "Station A" User Panel Operation

Depicted in figure 2 is a typical layout of "Station A" user panels. Note that "Station A" consist of three panels 1) Station A user panel, 2) Station A door panel, and 3) System Controller panel.

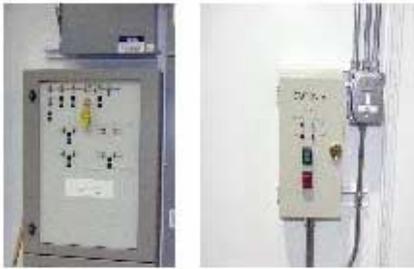

Fig. 2 PSS User Panels

## 3.0 TESTING METHODOLOGY

In this section, we describe how the PSS system requirements and validation procedures were structured using the Vali/Test Pro methodology. Note that the Vali/Test Pro is a methodology developed by Interim Technology Company and is widely used in the testing industry as a means to provide a visual approach to validating requirement coverage of a system or software application. This methodology provided us with the means to decompose PSS system requirement documents (SAD, DOE, ES&H) into their high, intermediate and detail levels. In addition, we decomposed the validation procedures for one PSS beamline into testing units called builds, test runs, and test cases (see figure 3). Afterwards, we mapped each requirement level to its respective testing unit in a validation matrix in order to measure the test coverage.

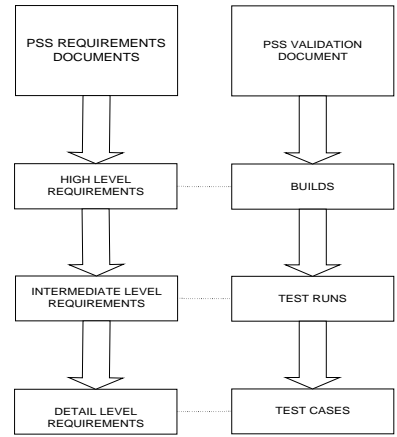

Fig. 3 Vali/Test Pro Testing Methodology

A validation matrix for each level of test requirements was constructed to map each level of requirements to its respective level of test. Below we define the different requirement levels and testing units used to defined each type of validation matrix.

- High level requirements: They represent the broadest categories of a system functions, such as business processes, including data entry and data capture, normal and exception processing, and database updates.

- Intermediate level requirements: These are components of the high-level requirements, often at the transaction or batch processing level. They may include sequences of actions to be taken, information to be displayed, error routines and associated messages.

- Detail Level Requirements: They are the specific steps in application processing, such as action steps field definitions, edit criteria, calculations, and error messages.

- Builds: A logically complete subset of an application that can be tested independently, and then integrated and tested with other builds.

- Test Runs: They consist of a set of related test cases that are used to validate the requirements at the intermediate level.

- Requirement Validation Matrix: A table that provides a cross-reference between a system's requirements and test case specifications.

## 4.0 AUTOMATED REAL-TIME TEST SYSTEM (ARTTS) DEVELOPMENT

In this section, we described how the ARTTS system was developed as a tool to automate the testing of the PSS

system described above. The development of the ARTTS system described here consists of three major areas of system integration: 1) Test Requirements, 2) Software Tools, and 3) System Hardware. Below we explain how the use of commercially available off-the-shelf software and hardware was integrated to perform all required testing for the PSS system.

### 4.1 Test Requirements

The Department of Energy regulations require the personnel safety system for each beamline to be validated every 6 months. In addition to the high level matrix described in section 3.0 there are an intermediate level and detail level matrices that were required to completely define the PSS system requirements.

### 4.2 Reasons to Automate PSS Testing

One reason to automate the PSS system testing is an increase in the need for additional validations over the past five years. Another reason to automate PSS testing is a steady increase in requests for software changes. Additional reasons to automate the PSS system are listed below.

- Manual testing of the PSS takes up to 3 days to validate one beamline depending on its configuration.

- An automated testing system will catch more software errors before final testing is performed on the lab floor.

- It will reduce the overall time required to validate a beamline.

- Since 98% of all software change requests (SCR's) are related to HMI and not safety, most of the testing performed on the lab floor can be moved to the ARTTS simulator. As a result, it will increase software reliability.

- It will reduce the cost of testing.

### 4.3 Software Tool Integration

Fully automated, the testing of a PSS system required the integration of four software tools onto a single computer platform along with a fifth tool that was set up to work on a separate notebook computer (see figure 8). Together the software tools provided the following functions: 1) Real time I/O simulation for programmable logic controllers, 2) Human Machine Interface (HMI) to user control panels, 3) Graphical User Interface (GUI) testing for Windows based applications, 4) Graphical test planning, batch testing and defect tracking, 5) PLC fault code verification for Allen Bradley PLC's, and 6) PLC fault code verification for GE PLC's.

The software tools selected to perform the above system functions were as follows:
- Programmable Industrial Control Simulation (PICS) software provided a real-time HMI and I/O simulator for testing PLC-based control system.
- WinRunner software provided a functional testing tool designed primarily to test graphical user interfaces of Windows base applications. It was used here in the ARTTS system to test and verify the state of the PSS system's I/O.
- Test Director software module provided graphical test planning, batch testing, defect tracking, and an interface to the WinRunner module.
- RSLogic is an A/B tool designed primarily to down load software to A/B's PLC's and to verify faults and task states in Chain B of the PSS system.
- State Logic is a GE software tool that provided the means to verify faults and task states in Chain A of the PSS system.

### 4.4 ARTTS Hardware

The ARTTS system's hardware is best described by the system hardware layout in figure 4. The following hardware makes up the ARTTS prototype system:

- Gateway GP7-500 – 500 MHZ CPU, 1GB RAM, 20 GB HD IDE, Vision Tek video card, A/B 1784KT card, AB-5236 SD, 10/100 MB/s Network Interface Cards

- Nokia 17 inch Monitor

- Allen Bradley PLC-5/30 Rack: A/B 120 VAC power supply, DH+A, DH+B, A/B PLC-5/30 CPU

- GE Fanuc Series 90-70 Rack: series 90-70 power supply, State Logic CPU, Ethernet interface

- Unicom Dyna-net/4 Null Hub

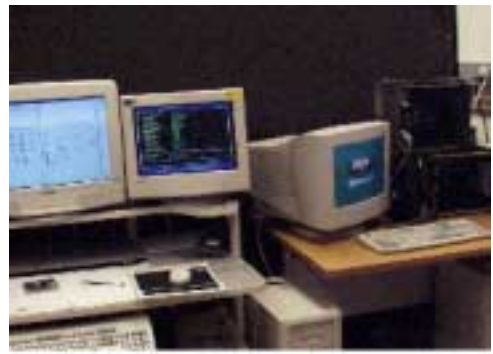

Fig. 4 ARTTS Hardware

### 4.5 ARTTS System Schematic

A schematic with all the ARTTS system components integrated together is shown in figure 5 below.

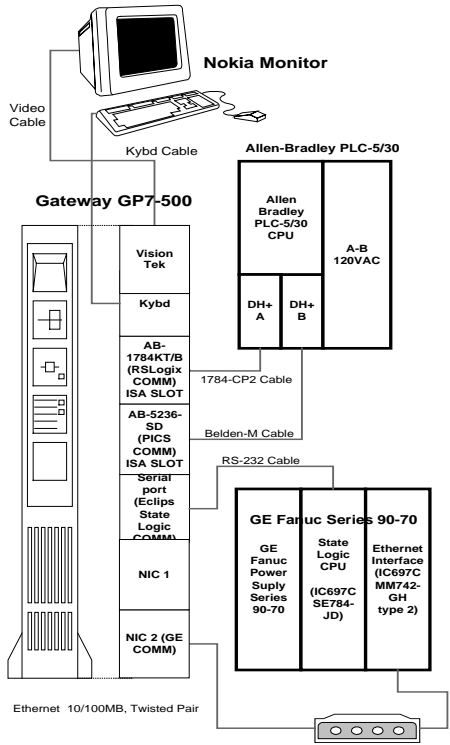

Fig. 5 ARTTS System Schematic

## 5.0 ARTTS OPERATION

### 5.1 Human Machine Interface (HMI)

A user interface to the ARTTS system is via a computer keyboard, mouse, and color screen computer monitor. Using PICS HMI graphical user interface, a user can control and monitor all functions of the PSS system by activating the proper switches on the user control panel and then monitoring the system function via LED outputs (see figure 6).

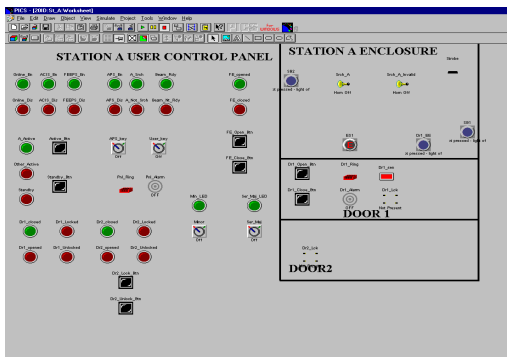

Fig. 6 ARTTS HMI Panel

*5.2 Test Planning*

A user can set up or activate batches of test to run by selecting test plan from TestDirector's menu. Once selected a batch of tests can be run by simply selecting Run Test from TestDirector's menu. This step automatically activates WinRunner to execute the selected test scripts (see figure 7).

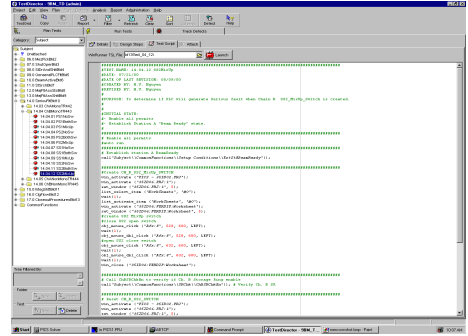

Fig. 7 ARTTS Test Planner

*5.3 Test Results*

Executed tests results are verified within Test Director by selecting the icon labeled "Details" from the menu.

The results of each test will be listed as pass or failed along with the date and time of execution of each test (see figure 8).

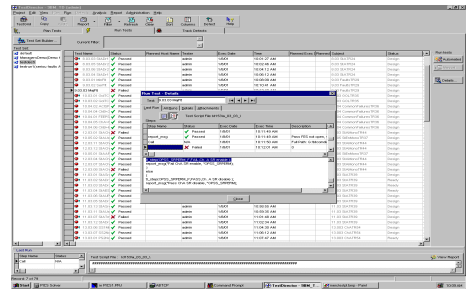

Fig. 8 ARTTS Test Results

### 6.0 ARTTS SYSTEM PERFORMANCE

The ARTTS system was very successful reducing the amount of time required to perform PSS testing. For example, the ARTTS system reduced the overall amount of time required to process a batch of 72 test cases from about 4 hours of manual testing to 1 hour and 36 minutes of automated testing. In general, the ARTTS system achieved an execution time of 1 minute and 20 seconds per test case. Furthermore, as a result of additional test coverage the ARTTS system greatly increased the reliability of the PSS software.

### 7.0 ARTTS SYSTEM ADVANTAGES VS DISADVANTAGES

*7.1 Advantages*

- Scalable.
- Major reduction in time required to validate a batch of test cases.
- Commercially available software.
- Commercially available hardware.
- Reduced development time.
- Easy to build.
- Better test coverage.

*7.2 Disadvantages*

- Uses 5 different software packages.
- Configuration management is more a challenge with multiple software packages

## BIOGRAPHIES:

Jon Hawkins, Group Leader
Argonne National Laboratory
Interlock Systems & Instrumentation Group
Bldg 401, Rm C1249
Argonne, IL 60439, USA
hawkins@aps.anl.gov

**Jon K. Hawkins** is a Group Leader at the Advanced Photon Source, Argonne National Laboratory. He is responsible for the design, testing, installation and validation of real-time embedded computer based mission critical systems. Mr. Hawkins is a senior engineer, whose experience ranges from high-speed analog and digital system design with full custom ASIC's to computer based system requirements/risk analysis. Mr. Hawkins holds a US patent, has been presented the Argonne National Laboratory Directors Award and has published numerous articles in technical journals.

Reginald B. Howard, P.E.
Argonne National Laboratory
Interlock Systems & Instrumentation Group
Bldg 401, Rm L1111
Argonne, IL 60439 USA
rhoward@aps.anl.gov

**Reginald B. Howard** is a test engineer consultant at Argonne National Laboratory while working for an IT consultant firm named Web Business Controls, Inc. He received his BS degree (1987) in electrical engineering from Old Dominion University. He has been working at Argonne National Laboratory for the last 2 years assisting with developing the automated real-time testing system for embedded control systems. In addition, he is responsible for developing and implementing a web based electronic documentation system for the Laboratory. Previously he worked as an electrical engineer with the Federal Aviation Administration (FAA) where he was responsible for developing the FAA's first touch screen control system for airport runway approach lights control systems at O'Hare International Airport Air Traffic Control Tower (ATCT). In addition, he has designed electrical distribution systems and lightning protection systems for ATCT facilities. Furthermore, he has worked in the manufacturing industry where he designed and tested electrical power conditioning products. Mr. Howard is a licensed professional engineer in the state of Wisconsin and is a member of the IEEE Computer Society and "The Instrumentation, Systems and Automation Society" (ISA).

Haung V. Nguyen
Argonne National Laboratory
Interlock Systems & Instrumentation Group
Bldg 401, Rm B1196
Argonne, IL 60439 USA
nguyen@aps.anl.gov

**Van Nguyen** is an Engineering Assistant at Argonne National Laboratory (ANL). He is also an undergrad computer science student who is planning to pursue a graduate degree in computer science at Illinois Institute of Technology (IIT). While working at ANL he has been responsible for testing Personnel Safety Systems (PSS) and responsible for maintaining Programmable Logic Controller (PLC) software for PSS systems. Furthermore he has been involved with developing and writing software tests scripts for the Automated Real-Time Test System (ARTTS).